# Discussion of various models related to cloud performance

Chaitanya Krishna Kande
Computer science department
Ohio State University
Columbus, Ohio, US
Kande.2@osu.edu

Christopher Stewart
Computer science department
Ohio State University
Columbus, Ohio, US
cstewart@cse.ohio-state.edu

## I. INTRODUCTION

Knowing the metrics related to cloud is very critical to enhance the performance of the cloud. To learn the metrics related to cloud like the infrastructure, structure and functionality of the cloud systems many models were described. This paper lists some of the technical models related to the cloud computing that enhances the cloud performance. The models that are discussed focusses on some metrics like page response time, admission control, enforcing elasticity to cloud services etc.

Chapter 1. K-Scope: Online Performance Tracking for Dynamic Cloud Applications

### A. Introduction

Cloud provides on-demand, flexible and easy-to-use resource provisioning. It is also an open platform where Cloud users can share software components, resources and services. These features gives rise to paradigms such as **Continuous Delivery**, where a Cloud application (e.g., Amazon web services) is delivered through frequent incremental updates and Shared **Platform Services,** which tells about platform-as-a -service which is shared by multiple applications in the cloud.

Present cloud services lack the ability to *continuously*, *efficiently* and *accurately* capture the most up to date performance characteristics of a dynamic Cloud application. Present modeling approaches are designed with a traditional static deployment in mind where an application runs on dedicated machines and its implementation does not change during the modeling process. Some of the approaches run offline while some cannot model multiple layer applications.

K-scope is a first online, multirequest, multi-layer application performance modeling approach. This model infers critical performance model metrics such as request service time at different functional layers which are basically unobservable. K – Scope also utilizes Kalman filters to continuously adjust model metrics to keep the model consistent with the dynamic Cloud application. This model can predict how much resources are need at different functional layers for a given performance.

Cloud application may have different functional layers which handle various requests. Queuing network model is the basic framework as it is general enough to model multi-layer multi-request applications. Kalman filter is a technique that estimates the changing performance characters.

### B. Queueing Network Model

These models capture the performance of complex computer systems at request level and system level.

For 3 class 2 tier system

$\lambda i$ = Arrival rate of class *i* jobs.
$Si\ j$ = Average service time of class *i* jobs at tier *j*.
$di$ = Additional delay for class *i* jobs in system.
$u0\ j$ = Background utilization for tier *j*.
$uj$ = Average utilization for tier *j*.
$Ri$ = Average response time for class *i* jobs in system

The system performance and resource utilization can be approximated by the queueing analytic relations below.

$uj = u0\ j + \lambda 1 S1\ j + \lambda 2 S2\ j + \lambda 3 S3\ j$, *j belongs to{1,2}*

$Ri = di + S_{i1}/1-u_1 + S_{i2}/1-u_2$ , i belongs to {1,2,3}.

The delay and service time parameters, however, are very difficult to measure directly.

In our 3-class 2-server example, the system parameters are

$x = (u_{01}, u_{02}, d1, d2, d3, S11, S21, S31, S12, S22, S32)^T$
X has to be estimated based on $Z = (u1,\ u2, R1, R2, R3)^T$
X can be estimated online using kalman Filter.

### C. Kalman Filter

Is commonly used to estimate the values of hidden state variables of a dynamic system that is excited by stochastic disturbances and stochastic measurement noise.

$x(t) = F(t)x(t-1) + w(t) = x(t-1) + w(t)$
$z(t) = H(t)x(t-1) + v(t)$.

*x*- *The state variable*
F(*t*) is the state transition model that describes the evolution of the state over time.
w(*t*) is the process noise (zero mean), multi-variant Normal distribution with certain covariance matrix
H(*t*) is the observation model
*v*(*t*) is the observation noise

Since the measurement model is a non-linear function of parameters, 'Extended' version of the Kalman filter is used.

The state of the filter is represented by two variables:
• ˆx(*t*/*t*) is the estimate of state at time *t* given observations up to and including time *t*.
• P(*t*/*t*) is the error covariance matrix (a quantitative measure of estimated accuracy of the state estimate

Predict:
ˆx(*t*/*t* −1) = F(*t*)ˆx(*t* −1/*t* −1)
P(*t*/*t* −1) = F(*t*)P(*t* −1/*t* −1)F*T* (*t*)+*Q*(*t*) , Q is covariance
Update:
H(*t*) =[δh/δx](ˆx(*t*/*t* −1)) , [δh/δx] a Jacobean
S(*t*) = H(*t*)P(*t*/*t* −1)H$^T$ (*t*)+*R*(*t*)
K(*t*) = P(*t*/*t* −1)H$^T$ (*t*)S−1(*t*)
ˆx(*t*/*t*) = ˆx(*t*/*t* −1)+K(*t*)(z(*t*)−h(ˆx(*t*/*t* −1)))(11)
P(*t*/*t*) = (I−K(*t*)H(*t*))P(*t*/*t* −1)

D. *Applications*

**Performance Diagnosis**: K-Scope explicitly estimates request service time at different layers, and
Provides a clear breakdown of the response time

**Answering What-If Queries** : A simple approach is that we first apply the model to track the system in a stable period; with all the model parameters estimated, the question can be generally solved by varying certain parameters and re-calculate the other parameters

**Capacity Planning.** K-Scope also simplifies capacity planning as application developers can leverage the performance model produced by K-Scope to virtually explore a large number of deployment options and predict the corresponding performance

CHAPTER 2. MAGPIE: ONLINE MODELLING AND PERFORMANCE-AWARE SYSTEMS

To understand the performance of distributed systems, the correlation of the interactions between components should be modelled. Magpie modelling service collates detailed traces from multiple machines in an e-commerce site, extracts request-specific audit trails, and constructs probabilistic models of request behavior. Even though this approach is promising, there are many challenges to building a truly ubiquitous, online modelling infrastructure.

A. *Introduction*

Computing at present totally depends on distributed architecture. It is easy to get the overall performance of the whole distributed system but is difficult to find the cause for worst performance of end-users. Aggregate statistics are insufficient to diagnose the problems at single users or systems. Accurate diagnosis requires detailed audit trail of each request and a model of normal request behavior. Comparing observed behavior with the model helps in finding anomalous requests and malfunctioning units. This should be the property for a basic operating system service. Each and every unit in the infrastructure should collect the performance traces and develop models which are available for online checking.

B. *Magpie, an online modelling infrastructure is based on two design principles*

Black-box instrumentation requires no source code modification to the measured system.

End-to-end *tracing* tracks not just aggregate statistics but each individual request's path through the system.

Even though fine grained, low overhead tracing exists, the challenge is a system that collects fine-grained traces from all software components; combines these traces across multiple machines; attributes trace events and resource usage to the initiating request; uses machine learning to build a probabilistic model of request behavior; and compares individual requests against this model to detect anomalies.

Performance debugging can be done using online monitoring and modelling. The potential uses of online monitoring and modelling include Capacity planning, Tracking workload level shifts, Detecting component failure, Comparison-based diagnosis, Bayesian Watchdogs, Monitoring SLAs etc.,

A feasible study is done for the above approach. i.e, is about tracing overheads, log data tractability and usefulness of the resulting model.

For this a e-commerce site a magpie (offline) demonstrator is constructed for various events which traces in-kernel activity, RPCs, system calls and network communication. Instrumentation points are created for various servers and they generate an event with timestamp.

Offline processing assembles logs from multiple machines, associates events with requests, and computes the resource usage of each request between successive events. To stitch together a request's control flow across multiple machines, we use the logged network send and receive events. Similarly, we could track requests across multiple thread pools on the web server, by instrumenting the thread synchronization primitives.

To estimate the overhead of logging, a test certain workload is carried out and inferred that the overhead is due to inefficient logging of system call.

C. Behavioral clustering

*For* the Indy performance prediction toolkit transactions are characterized by their URL but magpie characterizes them by behavior, which merges logs and serializes the requests events as event string .The event strings are clustered based on Euclidean distance. Behavioral clustering is better than URL approach.

*Behavioral* clustering alone is sufficient to identify the requests anomalies but to identify the events out of space probabilistic state machine is modelled. A probabilistic state machine can be represented as a stochastic context-free grammar (SCFG).

*Given* a set of example strings, the ALERGIA algorithm derives an SCFG in linear time by recursively merging similar portions of their prefix tree. The efficiency, combined with the enhanced information in the resulting model, has encouraged us to apply ALERGIA to Magpie request event strings.

D. Related Work

The closest relative to Magpie is Pinpoint which focus on fault detection rather than performance analysis. Scout and SEDA require explicitly defined paths along which requests travel through the system. In contrast, Magpie infers paths by combining event logs generated by black-box instrumentation. Whole Path Profiling traces program execution at the basic block level; Magpie's paths are at a much coarser granularity, but can span multiple machines.

Apart from some other techniques, few model checking approaches infer correctness models from source code analysis or runtime monitoring; this is similar to our approach of inferring performance models.

CHAPTER 3. Adaptive, Model-driven Auto scaling for Cloud Applications

Applications with dynamic workload demand access to flexible infrastructure for performance and to minimize costs. Even though cloud computing provides the required elasticity, cloud service providers lack access to user applications which makes difficult to scale the underlying infrastructure. This paper proposes a new cloud service Dependable Compute Cloud (DC2) that automatically scales the infrastructure for user requirements. DC2 employs kalman filtering to automatically learn the system parameters of applications and to proactively scale the infrastructure.

While cloud computing is promising, it is up to the application owner to take its advantages, i.e., the application owner has to decide how to scale the deployment for dynamic workload. Dynamically sizing a deployment is challenging because the application owner should have knowledge about the dynamics of the application and also should have expertise on when and how to resize the deployment.

DC2 leverages resource-level and application-level statistics to infer the underlying system parameters, to determine the scaling action. DC2 has a modelling and execution engine (kalman filtering) for this purpose. Kalman filtering provides estimates without accurate model. So queuing theory can be used as a model.

A. Implementation

The application owner has to provide the initial deployment model and the performance SLA requirements. The Application Deployer customizes the image for deployment and ties up the endpoints for the application during installation and configuration. Chef server automated the installation of software on VMs during boot. OpenStack is the operating system being used. The VMs for the application are created on an OpenStack managed private cloud deployment on Soft Layer. The Cloud Pool component is a logical entity that models the application and issues the directives (such as VM scale up/down) required to maintain the performance SLA for the application. The Monitoring Agent is responsible for retrieving the resource-level metrics from the hypervisor and the application-level metrics from the application. The Modeling + Optimization Engine takes as input the monitored metrics and outputs a list of directives indicating the addition or removal of VMs, migration of VMs, or a change in the resources allocated to VMs. These directives are passed on to the Policy-based Execution Engine that issues commands to OpenStack API, that in turn performs the scaling.

B. Modelling

The modeling engine lies at the heart of DC2 approach. Queueing-network model is used to approximate multi-tier cloud application. However, since cannot access the user application to derive the parameters of the model, Kalman filtering technique is used to infer these unobservable parameters.

Queuing-network model and Kalman filtering were described in the paper "K-Scope: Online Performance Tracking for Dynamic Cloud Applications".

The Kalman filtering technique gives us estimates of the unobservable system parameters. These estimates, along with queuing model is used to predict the future values.

It initially takes about a minute for estimates to converge. After convergence, the estimated values are in very good agreement with the monitored values, thus validating our technique and highlighting its accuracy.

The change in the workload causes a change in the service time of the requests. Kalman filter detects this change based on the monitored values, and quickly adapts its estimates to converge to the new system state. The estimated values of the system state are used to compute the required scaling actions for DC2.

*C. Related Work*

Auto-scaling approaches;

Prediction models and Control theoretic techniques rely on system pro- filing to convert the system state into scaling actions. Black box models do not need any application information for scaling. Grey-box approaches typically require less time to converge and infer the system state as opposed to black-box models.

Kalman filtering approaches and also Rule-based approaches are the ones that are related to Scaling of the cloud infrastructure.

# CHAPTER 4. ADVISE – a Framework for Evaluating Cloud Service Elasticity Behavior

Complex cloud services rely on different elasticity control processes to deal with dynamic requirement changes and workloads. However, enforcing elasticity control process does not lead to optimal gain due to the complexity of service structures, deployment strategies and underlying infrastructure dynamics. Therefore, being able, a priori, to estimate and evaluate the relation between cloud service elasticity behavior and elasticity control processes is crucial for runtime choices of appropriate elasticity control processes. This paper discuss about ADVISE, a framework evaluates a cloud service elasticity behavior by using some of clustering techniques.

**A common** approach used by many elasticity controllers is to monitor the cloud service and (de-)provision virtual instances when a metric threshold is violated. Other approaches based on service profiling or learning from historic information have been proposed. These evaluate only low-level VM metrics and not the elasticity decisions based on multiple levels.

This paper discusses ADVISE(evaluating cloud service elasticity behavior) framework estimates cloud service elasticity behavior by using details like service structure, deployment strategies, and underlying infrastructure dynamics. Core of ADVISE is a cluster based evaluation process to compute elasticity behavior.

*A. Cloud Service Structural and Runtime Information*

A cloud service composes of service topologies , a group of semantically connected service units or service parts. A service unit (e.g., a web service) represents a module offering computation or data capabilities .Service model consists of Structural Information, Infrastructure System Information and Elasticity Information.

Elasticity information consists of metrics, elasticity capabilities etc., being associated with the service parts. Elasticity Capabilities are grouped together as Elasticity Control Processes (ECP s) and inflict specific elasticity behaviors upon enforcement to different SP s, which are modelled as Service Part Behaviors. The behavior of a cloud service, denoted as BehaviorCloudService, over a period of time is defined as the set of all cloud service SP behaviors. The above information is captured and managed at runtime through an Elasticity Dependency Graph.

*B. Elasticity control processes*

Elasticity capabilities (ECs) are the set of actions associated with a cloud service, which a cloud service stakeholder (e.g., an elasticity controller) may invoke, and which affect the behavior of a cloud service.
Elasticity Control Processes (ECP) are sequences of elasticity capabilities $ECP_i = [EC_{i1} \rightarrow EC_{i2} \rightarrow ... \rightarrow EC_{in}]$, which can be abstracted into higher level capabilities having predictable effects on the cloud service. An ECP causes a change to the elasticity dependency graph and to the virtual infrastructure related information.

*C. Cloud Service Elasticity during Runtime*

To be able to estimate the effects of ECP s upon SP s, we rely on the elasticity dependency graph which captures all the variables that contribute to cloud service elasticity behavior evolution. During the deployment in the controllers learn about structural information and after deployment they learn about infrastructure info and elasticity information. This info is updated continuously at runtime.

Infrastructure resources, as mentioned previously, have associated elasticity capabilities, that describe the change(s) to be enforced and the mechanisms for triggering them.

*D. Evaluating Cloud Service Elasticity Behavior.*

It is learning the behavior of different cloud service parts, and their relation to different ECP s, not only with directly linked ones, and estimating the effect of an ECP, in time, considering the correlations among several metrics and among several service parts. The Learning Process used to determine cloud service part behavior is depicted.

*E. Learning Process.*

This takes the input as each metric's evolution over time .To estimate the expected evolution in metrics due to enforcing an ECP , for each monitored metric ,RTS (Relevant Time series Section) is selected and compared with previously encountered metrics evolution over time. RTS depends on avg. time needed to enforce ECP.

Clustering is used to detect the expected behavior due to enforcing an ECP.Clusters are constructed based on all SPs and each ECP based on distance between behavior points. The objective function of this process is finding the multi-dimensional behavior point, which minimizes the distance among points belonging to the same cluster .After obtaining the point clusters a correlation matrix is constructed for all SPs. This matrix is continuously updated when behavior points move from one cluster to another, or when new ECP s are enforced, thus, increasing the knowledge base.

*F. Determining the Expected elasticity behavior*

The latest metrics of each SP, change in the metric values and the ECP which is consider for enforcement is selected. The expected behavior consists of a tuple cluster centroids from clusters constructed during learning process and are closer to current metrics behaviors. The result is for each metric of SP , a list of expected values from the enforcement of ECP.

G. Experiments

To evaluate the above approach, an ADVISE framework is developed based on previous concepts. ADVISE gathers various information and a test bed with two cloud services is deployed on to the cloud. On these cloud services ECPs are enforced on various SPs.

An experiment for web application (3-tier) to evaluate the ADVISE is done by generating client requests at stable rate where load depends on size of the video and on the type of the requests. Load balancer tier is used for client requests distribution, application tier contains video streaming web service and A Cassandra NoSQL Distributed Data Storage Backend from where the necessary video content is retrieved.

The observed and the estimated behavior for the Application Server Tier of the cloud service when a remove application server from tier ECP occurs (ECP1).  At first, we observe that the average request throughput per application server is decreasing. This is due to two possible cases: (i) the video storage backend is under-provisioned and cannot satisfy the current number of requests which, in turn, results in requests being queued; (ii) there is a sudden drop in client requests which indicates that the application servers are not utilized efficiently. We observe that after the scale in action occurs, the average request throughput and busy thread number rises which denotes that this behavior corresponds to the second case where resources are now efficiently utilized. ADVISE revealed an insightful correlation between two metrics to consider when deciding which ECP to enforce for this behavior. Similarly, we can depict both the observed and the estimated behavior for the Distributed Video Storage Backend when a scale out action occurs (add Cassandra node to ring) due to high CPU utilization. We observe that after the scale out action occurs, the actual CPU utilization decreases to a normal value as also indicated by the estimation. Finally, we conclude that the ADVISE estimation successfully follows the actual behavior pattern and that in both cases, as time passes, the curves tend to converge. From the above experiment we can conclude that ADVISE framework indeed able to advise the elasticity controllers about cloud service to improve cloud service elasticity.

CHAPTER 5 .Connection and Performance Model Driven
Optimization of Page view Response Time

Connection and Performance Model Driven Optimization (CP-MDO), a novel approach for providing optimal QoS as defined by a cost objective based on client perceived page view response time and page view drop rate. This approach combines two models, latency model that captures interactions between browsers and servers across network protocol layers and a server performance model which that models performance across all the tiers. This approach solves the admission control problem.

The clients to the website are classified into different classes and are given admission(rejection rate per class) to the website based on class and also the response time is decided based upon the class. This approach formulates multiple class admission control problem as an optimization problem with a cost objective. There exists a complex relation between the admission control drop rate and the page view response time. The server may contain many tiers. When a client or browser sends a request it may be served by multiple layers where each layer respond differently under heavy loads and this contributes to differences in overall pageview latency, with one tier being identified as the latency bottleneck. The second contribution is an for addressing a key issue involved in determining the optimal admission control rate for multiple classes under heavy load.

## A. CONNECTION AND PAGEVIEW LATENCY MODEL

A Model is designed for the latency of page view download based on browser server complex interactions and the effects that admission control drops have on the client perceived latency. The interaction between the browser(client) and server happens in this manner. At first connection is established between them, then client sends a request and the response is an HTML object(page view) which may contain embedded objects for which client establishes parallel connections to download.

Many existing techniques for differentiating service rely on admission control as a form of load shedding to ensure that requests of low priority clients do not interfere with the response time perceived by high priority clients. However dropping connections have significant impact on interactions between client and server. In the case of TCP connection when client sends connection request and if it is dropped due to admission control, again the client sends the request after certain time, repeats this step until the connection is established or maximum number of requests reached. For admission control this property of TCP communications must be taken into account. The admission control must have a method for measuring latencies as seen by the client. An admission control may use server response time which is orders of magnitude different from client response time.

The page view latency model is based on all the above behaviors and the second connection is not established until the HTML page and its embedded objects are not retrieved for first connection.

Response time = Time for connection1 + time for getting HTML +max(time for getting embedded objects, time for connection2 + time embedded for 2).

The mean page view response time is avg, over all page view instances.

## B. SERVER PERFORMANCE MODEL

This model is about server complex performance using queuing theory. In this model page view download or response consists of server response time and service rate. This model does not take into account the communication establishment.

$T_s$ is the server latency in creating the HTML page (or embedded objects) and $T_d$ s the transfer latency. The page view response is $T_s+T_d$.

The server response time for a single server with a single service class under an open model of arrivals is given as

$T_s = (1/u)/(1-p)$, $p= x/u$ where x is admittance rate and p is the server utilization, u is service rate This can be extended for multitier multi class services.

$T_d$ depends on the time for HTML response and time for embedded objects. $T_d$ varies based on the size for embedded objects.

## C. CLIENT SERVER INTERACTION MODEL

It is a combination of pageview latency model and server performance model. Client perceives that the browser traverses through some intermediate nodes for single pageview download. The path taken by browser depends on server. The server maintains certain clients in each state and controls the transition rate between nodes through admission control and service rates. Each node has certain state with associated latency. States indicate the attempts made by certain browser to connect to the server and get the response pageview. States can be a processing of HTML page or processing of embedded objects in the HTML page or the termination state that corresponds to the success or failed termination of a pageview download. Arcs between nodes are labelled by traffic intensity; traffic intensity is defined in terms of acceptance rate of the connections and the drop probability.

From the perspective of single client the browser traverses a path through states and it may be processing more than one embedded objects by connecting in parallel to server and from server's perspective resources are consumed only when the client is processing HTML or embedded objects.This illustrates the two components of page view latency: TCP connection delays and server response time delay. This model captures that the established connection can be re used to download multiple embedded objects in the page view but the connection is terminated once the whole pageview is complete.

The traffic intensity throughout the model can be recursively expressed as a function of the external arrival rate, and the admission control drop probabilities.

## D. MINIMIZING CONNECTION TIME

Minimizing Connection Time over One Connection
For the pageviews with no embedded objects the latency of producing and transferring the HTML is same for all downloads. The optimal values for the admission control drop probabilities can be determined using the mean pageview response time. To minimize the response time the acceptance rate of the connection depends on the latency of the response. If the latency is small, acceptance is high. To determine the largest possible value for acceptance rate of particular connection server complex capacity should be enough to handle all the requests. If the server is capacity is sufficient then the optimal solution is to accept the first SYN for all incoming page requests else the optimal solution is to throttle pageview requests whose acceptancy rate is less than capacity. Connection latency for admitted pageview requests is minimized when only the first SYN request on the first connection is accepted.

Similarly for two connections, to minimize connection latency for admitted pageview requests, always accept the first SYN of the second connection. So we can say that for client server model we can state that if an admission control mechanism seeks to minimize the TCP connection establishment latency for admitted pageview requests, the optimal course of action is to accept both the first and second connection attempts when their initial SYN is received. An obvious corollary being, to deny a pageview request all SYNs on the first connection must be dropped. Any admission control mechanism that is unable to distinguish between the first and second connections of a pageview download or is unable to distinguish between a 1st, 2nd or 3rd SYN transmission is by definition sub-optimal and unable to manage client perceived pageview response time.

*E. MINIMIZING SERVER RESPONSE TIME AND PAGEVIEW DROP RATE*

A cost objective is proposed based on server response time and pageview drop rate. This objective function is based on class pageview acceptance rate indicating there exists an optimal acceptance rate vector, opt accept, that minimizes the cost objective. The mean response time of a class (requests of a particular class), its pageview drop rate and the relative weight importance of a class can be applied to minimize the cost objective i.e, an admission control mechanism must be developed that is capable of enforcing the optimal per class acceptance rates. Such a mechanism must categorize clients into service classes and measure the response time as perceived by the remote client. Observed Inter-Arrival Time (IAT Control) algorithm enforces the above solution.